\documentclass[letterpaper]{article} 
\usepackage{aaai2027}  
\usepackage[hyphens]{url}  
\usepackage{graphicx} 
\usepackage{natbib}  
\usepackage{caption} 
\usepackage{algorithm}
\usepackage{algorithmic}

\usepackage{newfloat}
\usepackage{listings}
\DeclareCaptionStyle{ruled}{labelfont=normalfont,labelsep=colon,strut=off} 
\floatstyle{ruled}
\newfloat{listing}{tb}{lst}{}
\floatname{listing}{Listing}

\usepackage{booktabs}

\usepackage{amsmath,amssymb,amsfonts}%
\usepackage{amsthm}%
\usepackage{mathrsfs}%

\title{The Accuracy Trap: Structural Scarcity Amplifies Relative Inequality in Algorithmic Allocation}
\author{
    Erina Seh-Young Moon\textsuperscript{\rm 1}\corresponding,\\
    Matthew Tamura\textsuperscript{\rm 1},
    Shion Guha\textsuperscript{\rm 1}\\}
\affiliations{
    \textsuperscript{\rm 1}University of Toronto\\

}

\begin{document}

\maketitle

\begin{abstract}
Algorithmic systems increasingly rank individuals for access to scarce public resources, from child welfare interventions to cancer treatment referrals. The prevailing fairness frame treats disparity as a property of biased data or deficient models, with remedies through calibration and debiasing. Under structural scarcity, where demand exceeds supply by an order of magnitude, allocation becomes a rationing problem, and the statistical properties of ranking diverge sharply from those of classification. We derive a scaling law $D \propto \exp(t \cdot \rho \cdot \Delta)$, in which relative disparity between two groups separated by a structural gap $\Delta$ grows in the product of the scarcity-induced threshold $t$ and rank-discrimination fidelity $\rho$. Scarcity and accuracy interact multiplicatively, producing exponentially larger between-group disparities. We term this dynamic the Accuracy Trap. We validate this Accuracy Trap through Monte Carlo simulation and two independent public-sector systems in Canadian child welfare and U.S. cancer care. Debiasing alone cannot dissolve the trap.
\end{abstract}


\section{Introduction}

The dominant paradigm in algorithmic governance treats inequality in automated decision-making as a product of biased training data or deficient model design \cite{barocas2016big, buolamwini2018gender, chouldechova2017fair}.
~This diagnosis has produced literature on debiasing techniques, fairness constraints, and impossibility theorems \cite{dwork2012fairness, hardt2016equality, kleinberg2016inherent, kleinberg2018algorithmic}.
~Recent work has questioned the normative legitimacy of optimising for predictive accuracy in social domains \cite{wang2024against}
~and argued that fairness must be understood through the lens of power and resource distribution rather than statistical constraints alone \cite{kasy2021fairness}.
~A large body of fairness literature and its critics share a common analytical frame. They evaluate algorithmic systems under implicit assumptions of resource sufficiency, asking whether the algorithm sorts people correctly rather than asking what happens when there is almost nothing to sort them into.

We argue that this frame overlooks the dominant constraint of the public-sector domain in which algorithmic allocation is most consequential, namely structural scarcity. In homelessness services, child welfare, and cancer treatment, demand routinely exceeds supply by an order of magnitude \cite{shafir2013scarcity, eubanks2018automating}. Even when bias has been identified in such systems, as in the widely cited case of racial 
~disparities in healthcare risk prediction \cite{obermeyer2019dissecting},
~the corrective focus remains on the algorithm's internal calibration rather than on the structural interaction between prediction accuracy and resource scarcity. Under these conditions, allocation ceases to be a matching problem and becomes a rationing problem. The system must rank individuals and then impose a threshold that excludes the vast majority. The statistical properties of ranking under such thresholds diverge sharply from those of classification, because selection occurs in the extreme tail of the score distribution, where the geometry of the problem is governed by tail behaviour rather than by mean accuracy.

The intuition for our results can be stated plainly. An imprecise ranking system, by virtue of its noise, randomises the selection of individuals near the threshold. This randomisation inadvertently preserves a nonzero probability that members of the group structurally disadvantaged by the allocation mechanism will be chosen. Here, we define structural disadvantage by the operational setup of the allocation system, which may differ or run counter from historical definitions of structural marginalization. As the system becomes more precise, the randomisation vanishes, and selection at the tail collapses onto whichever group sits closer to the threshold in expectation. Imprecision, in other words, has been operating as an unintentional equity buffer all along. We formalise this dynamic and term this the Accuracy Trap. We derive the rate at which the buffer erodes as fidelity rises, and show that it produces a regime of near-total exclusion at the tail. The Trap is agnostic to the normative value of the resource being allocated. Whether the amplified disparity benefits or harms specific populations depends on factors external to the model, i.e., whether the allocated resource is a benefit or a burden, and whether that population sits above or below the threshold. 


Our work demonstrates that under scarcity, algorithmic sorting is precisely such a mechanism, and we derive the rate at which it operates. When a ranking algorithm with rank-discrimination fidelity~$\rho$ selects from a population in which an advantaged group (as operationalised by the allocation system) is separated from a disadvantaged group by a structural gap~$\Delta$, the relative disparity in selection probability grows as $D \propto \exp(t \cdot \rho \cdot \Delta)$, where $t$ is the scarcity-induced selection threshold. Scarcity and accuracy interact multiplicatively. Modest improvements in algorithmic precision that are inconsequential under abundance produce exponentially larger disparities under scarcity, thus giving rise to the Accuracy Trap. Model fidelity ($\rho$) amplifies the structural gap ($\Delta$) encoded by the allocation system. 

The structural gap $\Delta$ in our model is exogenous, but it is not a primitive of the social world. $\Delta$ encodes the residue of historical and structural inequality that remains after upstream debiasing has done its work. Even when training data has been corrected, calibration achieved across protected groups, and disparate-impact constraints satisfied, the residual $\Delta$ that remains will be amplified exponentially by the interaction of scarcity and accuracy. Debiasing the input does not dissolve the trap. It merely sets the value of $\Delta$ at which the trap operates. 

This reframing has consequences for how we think about bureaucratic discretion. A substantial literature has documented the harms of street-level discretion, including racial bias in caseworker decisions \cite{eubanks2018automating, lipsky2010street, saxena2023rethinking}.
~We do not dispute these findings. We show, alongside them, that discretionary noise has a structural side-effect operating independently of its case-level harms. By introducing variance into the ranking function, imprecision prevents the system from fully resolving the distributional gap between groups, and maintains nonzero selection probabilities for groups positioned unfavorably by the allocation system. Prior ethnographic work has documented this dynamic at the case level \cite{saxena2021framework, akpinar2021effect}. 


Our contribution is to formalise the mechanism and quantify the rate at which the buffer erodes (Section \ref{sec:sec2}). Moon and Guha \shortcite{moon2026paradox} presents a version of the phenomenon in a discrete combinatorial framework on finite allocation instances. This work substantially extends that result by deriving the continuous asymptotic scaling law in Gaussian and log-normal regimes. We validate our findings through Monte Carlo simulation and empirical analysis of two independent public-sector allocation systems operating under scarcity (Section \ref{sec:empirical_validation}). Despite differences in country, population, institutional context, and algorithmic architecture, the systems exhibit disparity patterns consistent with the theoretical prediction.

\subsection{Related Work}

Sociological theory has long recognised that institutions amplify categorical inequality. Tilly \shortcite{tilly1998durable} formalised how organisational mechanisms transform small between-group differences into durable advantages, and Merton \shortcite{merton1968matthew} described how initial advantages compound through cumulative processes. When automated decision-making systems are deployed in such institutions, Liu et al. \shortcite{Liu_Barocas_Kleinberg_Levy_2024} demonstrate that the predictive accuracy of these algorithms alone cannot guarantee better outcomes; the context in which they are implemented is equally important in determining the success of these tools. Despite the critical role of the deployment context, much of the literature on algorithms deployed in institutional settings has largely focused on examining the model itself. Studies have examined the feasibility of deploying risk-based allocation algorithms for resource-scarce domains \cite{preventing_eviction24} and explored which allocation objectives should be optimized in resource allocation algorithms \cite{Kube_Das_Fowler_2019}. An extensive body of literature has also introduced fairness metrics to assess these tools \cite{Li_Wu_Su_2023, barocas-hardt-narayanan} and investigated the (in)compatibility of different fairness criteria \cite{Rosenblatt_Witter_2023, chouldechova2017fair, kleinberg2016inherent, impossibility21, mashiat2022}. This study is related to prior literature that ask how fairness criteria can be satisfied under capacity constraints \cite{jo2023fairness, Nguyen_Das_Garnett_2021, mashiat2022, Dong_Garg_Dean_2026}. However, whereas those studies evaluate fairness under a fixed model, we investigate how varying model fidelity affects allocation disparities under resource scarcity.  Our work is also related to studies that examine the role of discretion and arbitrariness in decision-making algorithms \cite{pokharel2024discretionary, saxena2021framework, Cooper_arbitrariness_2024}.

\section{Model} \label{sec:sec2}

\subsection{Formal Derivation of the Accuracy Trap} \label{sec:sec2.1}

We formalize the Accuracy Trap by first presenting the model under a Gaussian regime. We model an allocation algorithm as a ranking function in which resources are assigned according to rank (e.g., those of higher rank or risk score receive resources first). 

We model a population partitioned into two groups, Group $A$ and Group $B$. Each individual has a latent risk $Z$, where $Z_\text{A} \sim \mathcal{N}(\mu_\text{A}, 1)$ and $Z_\text{B} \sim \mathcal{N}(\mu_\text{B}, 1)$. We define the structural gap $\Delta = \mu_\text{A} - \mu_\text{B} > 0$. In practice, $Z$ is often unknown. Decision-makers instead observe a score, $Y$ that is subject to a fidelity parameter, $\rho \in [0,1]$, and independent noise $\varepsilon \sim \mathcal{N}(0,1)$: 
\begin{equation} \label{eq:y_equation}
Y = \rho Z + \sqrt{1 - \rho^2}\,\varepsilon, \qquad \varepsilon \perp Z
\end{equation}

\noindent\ We interpret $\rho$ as a rank-discrimination parameter rather than a calibration parameter, since threshold-based selection is sensitive only to the relative ordering of scores. By construction, Equation \ref{eq:y_equation} preserves unit variance such that $Y_\text{g} \sim \mathcal{N}(\rho\mu_\text{g}, 1)$. 

Following the allocation of resources under this algorithmic system, we define relative disparity between groups as the ratio of selection probabilities between the two groups:
\begin{equation}
D(t,\rho) = \frac{P(Y_\text{A} > t)}{P(Y_\text{B} > t)}
\end{equation}

\noindent\ Here, $t$ is the scarcity-induced selection threshold. Denoting $\sigma$ as the fraction of the population that can be selected to receive resources, as resources become increasingly scarce ($\sigma \to 0$), this threshold moves into the extreme right tail ($t \to \infty$). Using Mill's ratio \cite{mills_1926, FROM2020123872}, we find that the dominant behavior of the relative disparity ratio is (see Appendix A.2 for a detailed derivation):
\begin{equation} \label{eq:log_relative_disp}
    \ln D(t,\rho) = t\rho\Delta + O(1) \quad \text{as } t \to \infty
\end{equation}

\noindent\ Thus, log relative disparity grows linearly in $t$ and $\rho$. Equivalently, exponentiating yields:
\begin{equation} \label{eq:relative_disp}
    D(t,\rho) \propto \exp(t \cdot \rho \cdot \Delta) \quad \text{as } t \to \infty
\end{equation}

\noindent\ Thus, we find that relative disparity $D$ scales exponentially. Taking the partial derivative of $D$ with respect to $\rho$ shows (see Appendix A.3. for a detailed derivation): 
\begin{equation} \label{eq:sensitivity_deriv}
    \frac{\partial D}{\partial \rho} \approx t\Delta \cdot D(t,\rho) \approx t\Delta \cdot \exp(t\rho\Delta)
\end{equation}

\noindent\ We, therefore, find scarcity amplifies sensitivity to relative disparity. When high resource scarcity pushes $t$ into the tail, the same marginal improvement in fidelity ($\rho$) produces a much larger multiplicative change in disparity, implying noise ($\rho$) has a structural effect. 

\subsection{The Accuracy Trap in Log-Normal Regimes} \label{sec:acctrap_lognormal}

To demonstrate the scale invariance of our findings in Section \ref{sec:sec2.1}, we now examine relative disparity within a log-normal framework. In this setting, we define the risk score $X = exp(Z)$, where $Z =\ln(X) \sim \mathcal{N}(\mu, 1)$. A selection threshold applied to the score ($X>T$) is equivalent to a threshold on the log-transformed risk scale $Z > \tau$, where $\tau = \ln(T)$. Substituting this relationship into Equation \ref{eq:relative_disp}, we find that in heavy-tailed regimes the relative disparity $D$ follows a power-law relationship (see Appendix A.4 for a formal derivation):
\begin{equation} \label{eq:logn_relative_disp}
D(T, \rho) \propto \exp( \ln(T) \cdot \rho\Delta ) \implies D(T, \rho) \propto T^{\rho\Delta}
 \end{equation}
 
\noindent\ These findings highlight that while the functional form of relative disparity changes under log-normal regimes, the core dynamics remain: increasing scarcity (i.e., raising $T$) or increasing model fidelity ($\rho$) monotonically amplifies relative disparity between groups.

\section{Empirical Validation} \label{sec:empirical_validation}
We empirically validate our theoretical findings from the previous section across three distinct settings: Monte Carlo simulations, healthcare, and child welfare. For the healthcare and child welfare validations, we operationalise the fidelity parameter $\rho$ through post-hoc noise injection around each system's current operating point, following $Y = \rho R + \sqrt{1-\rho^2}\epsilon$ where $R$ is the deployed risk score and $\epsilon$ is independent Gaussian noise. This procedure probes the allocative volatility of each system as a function of its ranking precision, and is not intended to simulate deployment of a modified algorithm. In each domain, we evaluate allocation disparities between population groups, with domain-specific group definitions detailed in each subsection below.

\subsection{Monte Carlo Simulation}

For our Monte Carlo simulations, we generated a synthetic population of 20,000~individuals, split equally into two groups: Group A ($n_\text{A} = 10,000$) and Group B ($n_\text{B} = 10,000$). We sampled a latent risk score ($z$) for the population, drawing from group-specific normal distributions with equal standard deviations ($s_\text{A} = s_\text{B} = 1$) but different means ($\mu_\text{A} = 0.8, \mu_\text{B} = 0.3$). We generated predicted risk scores ($Y$) by adding Gaussian noise ($\varepsilon \sim \mathcal{N}(0, 1)$) and applying a fidelity parameter ($\rho$) ranging from 0.2 to 1 onto the latent risk values following $Y = \rho z + \sqrt{1 - \rho^2}\varepsilon$. We varied resource scarcity levels where individuals were selected to receive a resource if their predicted score $Y$ exceeded the $(1-\sigma)$-th percentile. To ensure robustness, we performed 25 iterations for each combination of $\sigma$ and $\rho$ values to calculate the log relative disparity between the two groups, 
$\ln(D) = \ln [P(Y \geq t \mid A) / P(Y \geq t \mid B)]$. In Supplementary Materials figures 2 and 4, we show further explorations of the Accuracy Trap under simulation as we vary inter-group mean, fidelity, and variance differences

\subsection{Child Welfare}
\subsubsection{Motivating Context for Child Welfare Empirical Validation}
Child welfare systems have begun exploring the use of LLMs on case note narrative data to identify families at risk of deviating from intended service goals  \cite{moon2026promises}. Defining a child welfare `case' as the family unit engaged with an agency, we conceptualised case-level `risk' as a measurable deviation from formalised service goals established for that family. Documented lack of progress on case goals can signal that safety concerns remain unaddressed and require further attention. This definition of risk enables the targeted allocation of scarce resources, such as staff time and support services, to closely support these cases. 

We used this risk metric to evaluate the log relative disparity between cases serviced by two distinct geographic clusters at a Canadian child welfare agency, Children's Aid Society Toronto (CAST): CAST's Inner Toronto teams (the urban core area) and Outer Toronto teams. The City of Toronto is a deeply fragmented city with high concentrations of poverty in pockets across the City \cite{income_ineq}. These heterogeneous socioeconomic factors have contributed to differences in the types of cases the agency needs to address. Following Pollock et al. \shortcite{Pollock2024}, who identified location as a significant predictor of child welfare service variation, we treated between-region differences as a detectable empirical signature of these structural differences. We tested whether Toronto's geographic socioeconomic differences drive measurable variations in case trajectories, which in turn, shape how algorithmic risk scoring interacts with allocation under scarcity. The Inner-versus-Outer Toronto contrast serves as an empirical test bed for the Accuracy Trap, with regional differences acting as a proxy for the structural gap ($\Delta$) in our formal model. 

\subsubsection{Dataset and Risk Metric Construction}
\begin{figure}
	\centering
	\includegraphics[width=0.48\textwidth]{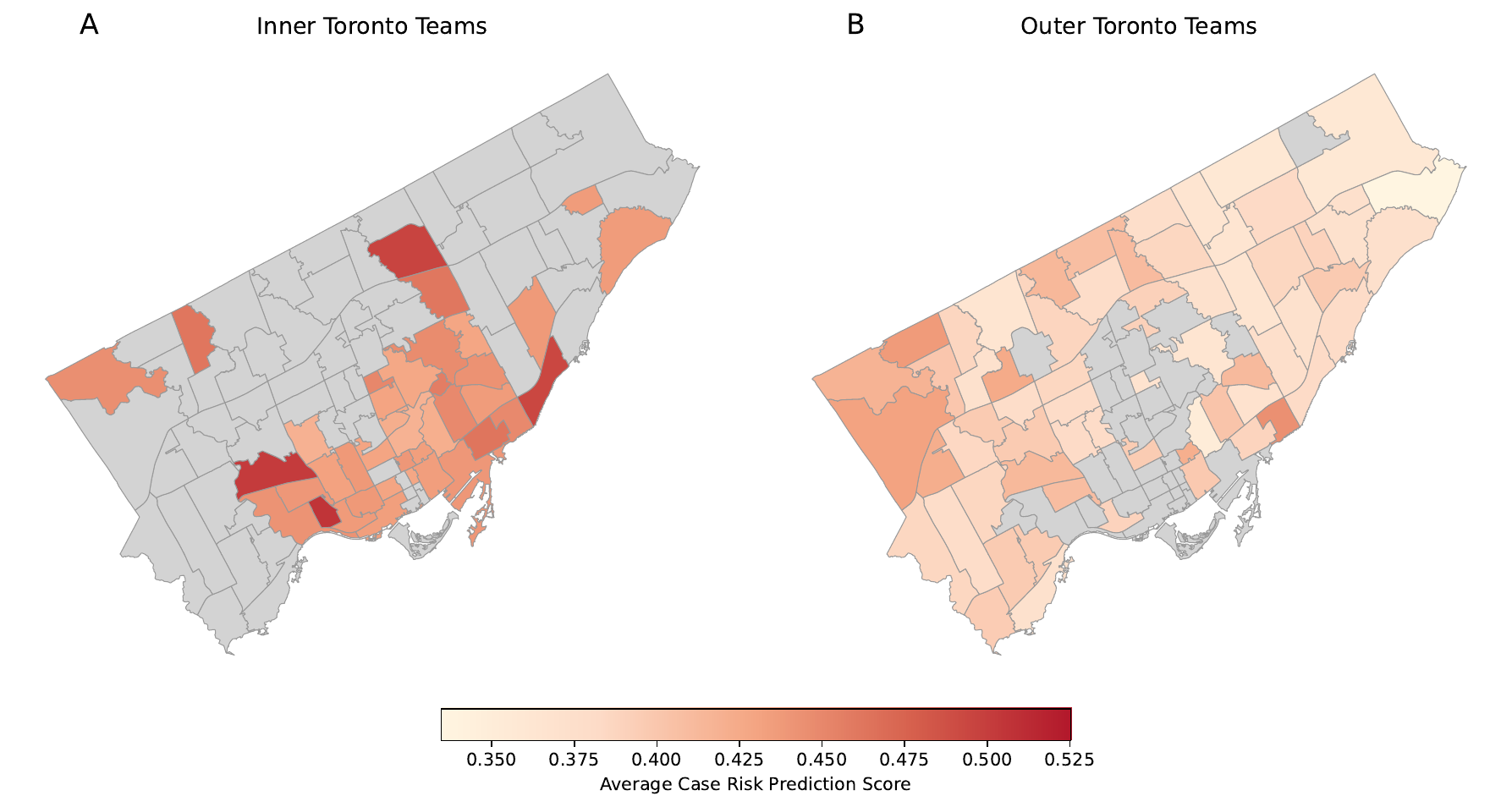} 

	\caption{Heatmap of mean case risk score ($R_\text{llm}$) across Toronto municipal regions. Gray regions indicate areas with no sampled cases. Higher risk scores (dark red) are more heavily concentrated among families served by Inner Toronto teams (\textbf{A}) compared to families served by Outer Toronto teams (\textbf{B}). These observed regional differences in $R_\text{llm}$ serve as our proxy for the structural gap $\Delta$ in our allocation model \cite{Pollock2024}. Boundaries are based on 2021 Census files from Statistics Canada \cite{statcan2021boundarymap}.}
	\label{fig:heatmap1} 
\end{figure}

We used a child welfare dataset comprising $n=37,201$~narrative case notes for $N=583$~families (i.e., `cases') collected during 2022-2025 in CAST. We obtained approval from the Research Ethics Board (REB) from our research institution to use this dataset.  To operationalise the risk metric, we used a Local LLM (Meta-Llama-3.1-8B), to classify whether each case note documented progress toward case goals ($1 = \text{aligned}$, $0 = \text{not aligned}$). We then computed a family-level risk score, $R_{\text{llm}}$, as one minus the proportion of goal-aligned case notes across all casenotes for that family. These risk scores ranged from 0 to 1, with higher values indicating greater goal deviation (higher risk). As shown in the geographic heatmap (Fig. \ref{fig:heatmap1}), average $R_{\text{llm}}$ risk scores vary significantly across regions. The mean for families served by Inner Toronto teams was 0.443 (95\% CI [0.408, 0.478]), compared with 0.385 (95\% CI [0.363, 0.406]) for families served by Outer Toronto teams, yielding an empirical structural gap of $\Delta = 0.058$. Child welfare professionals and researchers previously validated the Local LLM’s classification performance in a prior study \cite{moon2026promises} by manually evaluating the LLM’s labels for a subset of the dataset’s case note records ($n=6,031$) (Supplementary Materials table 1 and section C.1. provides further details).

\subsubsection{Disparity Analysis Strategy}
We compared the log relative disparity between cases managed by the two teams under varying ($\sigma$) and fidelity ($\rho$) levels. We defined log relative disparity as $\ln(D) = \ln \left( \frac{P(\text{Selected} \mid \text{Inner Toronto})}{P(\text{Selected} \mid \text{Outer Toronto})} \right)$. To examine the effects of varying fidelity, we applied a post hoc noise-injection approach. We applied the fidelity parameter to the predicted risk score such that $Y=\rho R_{\text{llm}} + \sqrt{1 - \rho^2}\,\varepsilon, \qquad \varepsilon \perp R_\text{llm}$. 
This design isolates the channel through which fidelity improvements would operate and a $\rho$ value of $1$ represents the deployed model's operating point. To ensure robust estimates of relative disparity, we employed a bootstrapping procedure, resampling with replacement for 1,000 iterations of the post hoc noise-injection approach. 


\subsection{Breast Cancer Prediction}
Delays in receiving breast cancer treatments can have critical implications for patient mortality rates, and resource-scarce conditions can exacerbate such delays \cite{Hanna_cancerdelay20}. Using data from the Surveillance, Epidemiology, and End Results (SEER) program, which provides detailed information on cancer incidence and survival in the United States, we developed a predictive risk model using a gradient boosting-based prediction model to estimate a risk score ($R_{\text{bcss}}$) for breast cancer-specific 5-year mortality. These risk scores ranged from 0 to 1, with higher values indicating greater mortality risk. For this empirical validation, we took a subset of the data from the SEER database to construct a dataset, comprised of cases of breast cancer diagnosed between 2012 and 2021. We examined relative log ratio disparity between Non-Hispanic White ($n=118,662$) and Non-Hispanic Black ($n=16,820$) population groups. The mean predicted 5-year mortality risk was 0.325 (95\% CI [0.324, 0.327]) for Non-Hispanic White patients and 0.418 (95\% CI [0.414, 0.422]) for Non-Hispanic Black patients, yielding an empirical structural gap of $\Delta = 0.093$ (Supplementary Materials table 2 and section C.2. provides further details). We defined log relative disparity as $\ln(D) = \ln \left( \frac{P(\text{Selected} \mid \text{Black})}{P(\text{Selected} \mid \text{White})} \right)$. We applied a post-hoc noise injection approach such that $Y = \rho R_{\text{bcss}} + \sqrt{1 - \rho^2}\,\varepsilon, \qquad \varepsilon \perp R_{\text{bcss}}$. 
To ensure robust estimates of log relative disparity, we repeated the allocation procedure across 25 independent simulations for combinations of fidelity ($\rho$) and resource scarcity levels ($\sigma$)

\subsection{Empirical Results} 

\begin{figure*} 
	\centering
	\includegraphics[width=0.5\textwidth]{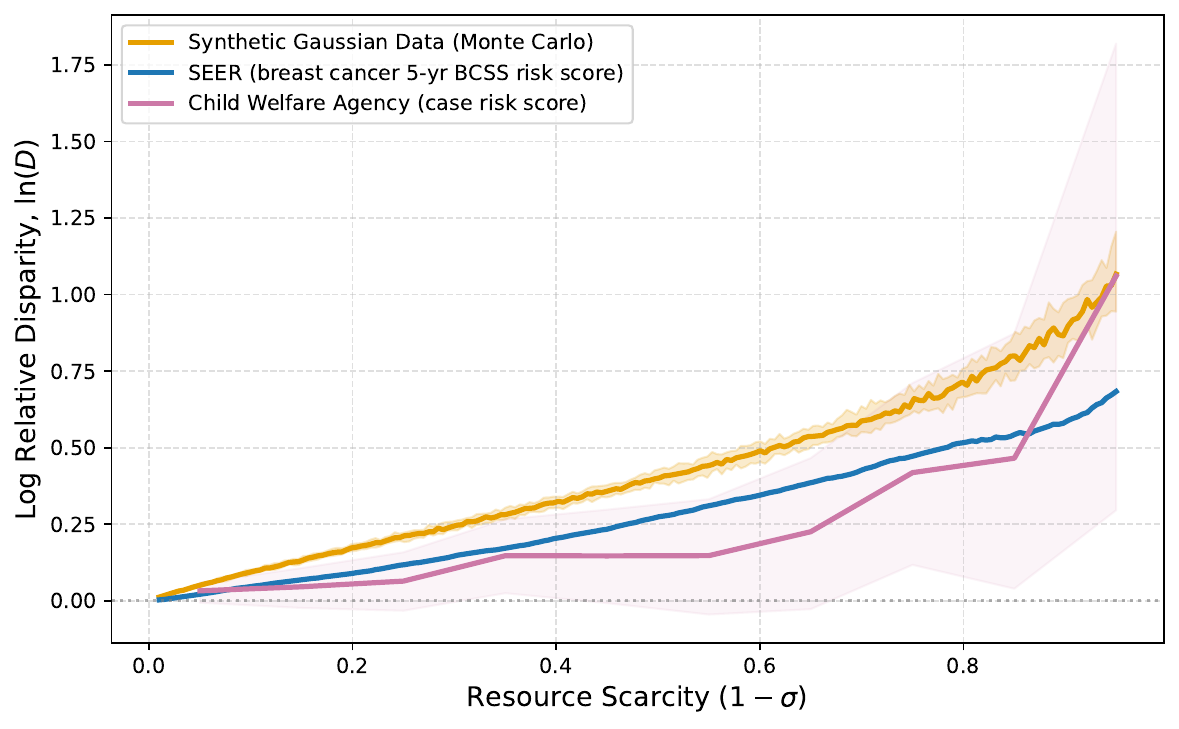} 

	\caption{Resource scarcity amplifies relative disparity in allocation. Plot shows median log relative disparity $\ln(D)$ as a function of resource scarcity (1--$\sigma$) at fidelity $\rho$ = 1 (model's deployed operating point). This pattern is observed across three domains: (yellow line) Monte Carlo simulation; (blue line) a gradient boosting risk model predicting 5-year breast cancer–specific survival (BCSS) using SEER registry data; and (pink line) an LLM-derived risk model measuring case progress divergence in child welfare data. Shaded regions denote 95\% confidence intervals.}
	\label{fig:summary1} 
\end{figure*}

Figure \ref{fig:summary1} shows that across all three settings, as resource scarcity increases ($1-\sigma \to 1$), disparities between population groups increase. The Monte Carlo simulation using synthetic Gaussian data (yellow line) provides the closest portrayal of the formalised Accuracy Trap, with narrow confidence interval bounds (shaded regions around the line) due to the idealised Gaussian nature of the synthetic data. The empirical results from the SEER cancer registry data (blue line) and child welfare data (pink line) closely mirror the simulation’s upward linear trajectory, providing real-world validation of the theoretical findings. The child welfare setting exhibits greater variability, with wider confidence intervals and more pronounced fluctuations, especially as resource scarcity intensifies. This likely stems from the smaller sample sizes in the extreme tail for the child welfare distribution compared to the other settings, which have larger datasets. Nonetheless, despite differences in variability between the three domains, we observe the monotonic increase in relative disparity.

\begin{figure*} 
	\centering
	\includegraphics[width=0.9\textwidth]{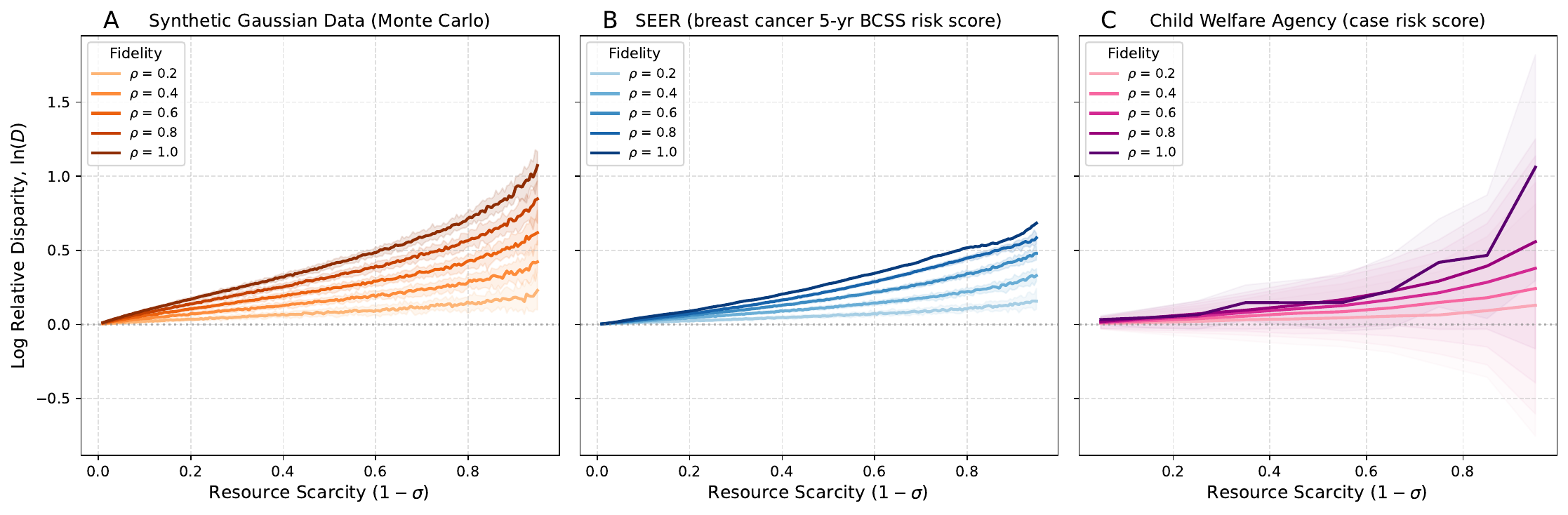}

	\caption{Model fidelity amplifies scarcity-driven relative disparity across three domains. (\textbf{A}) synthetic Monte Carlo generated data samples drawn from Gaussian distributions $\Delta = 0.5$, $sd = 1$, (\textbf{B}) Gradient Boosting model generating BCSS risk prediction scores trained on SEER Breast Cancer data, and (\textbf{C}) LLM-risk score generating algorithm applied on Canadian Child Welfare Agency casenote data. Across the three domains, low-fidelity models (lighter-colored lines) attenuate the impact of scarcity, while higher-fidelity models (darker-colored lines) increase structural gaps at scarcity. The shaded regions denote 95\% confidence intervals and show that uncertainty increases with greater scarcity. Lines depict median log relative disparity.}
	\label{fig:facet1} 
\end{figure*}

We next consider how the disparity changes as model fidelity ($\rho$) varies. Figure \ref{fig:facet1} shows that increasing model fidelity exacerbates disparities between population groups. While a lower-fidelity model (lighter-colored lines) produces relatively flat disparity trajectories even when resources become limited, higher-fidelity models (darker lines) yield stark relative log disparities. This figure provides empirical evidence of the sensitivity results depicted in Equation \ref{eq:sensitivity_deriv}, confirming that the slope of the Accuracy Trap is affected by fidelity ($\rho$). As $\rho$ increases, the allocation algorithm becomes better at reflecting the risk $Z$, thereby translating the underlying structural gap $\Delta$ into acute relative disparities in high-scarcity regimes (see also figure 3 in the Supplementary Materials). This suggests a paradoxical trade-off emerges in resource-scarce environments. The more accurately a model ranks individuals per the intended allocation mechanism, the more it sharply excludes groups sitting below the threshold \cite{shomik24}. 

\section{Discussion}
Our results demonstrate that in allocation systems operating under structural scarcity, improving algorithmic fidelity amplifies relative inequality between groups. This finding holds across simulation, two empirically independent public-sector domains, and two distributional regimes (Gaussian and log-normal). The scaling relationship $D \propto \exp(t \cdot \rho \cdot \Delta)$ implies that scarcity and accuracy are not independent policy variables. They interact multiplicatively, producing disparities that grow exponentially as both increase.

We read this result as exposing a gap in how the algorithmic fairness field has framed its central problem. Much of prior literature has spent over a decade developing classifier-level fairness constraints under implicit assumptions of resource sufficiency. However these constraints do not bind in actual public-sector deployments, which operate under rationing. A fairness intervention that equalises false-positive rates at an operating point in the body of the score distribution has nothing to say about allocative volatility in the tail, where rationing actually occurs. Recent work has begun to study model fairness under scarcity constraints, but has often treated the model as operating at a fixed point \cite{jo2023fairness, mashiat2022}. The Accuracy Trap builds on this literature by examining how allocation disparities evolve as model fidelity changes under structural scarcity. 


The Accuracy Trap alters how we should understand imprecision in public administration. Bureaucratic discretion has traditionally been characterised as inefficiency, and the move toward structured instruments has been justified as a corrective. Our results suggest that the noise eliminated by structured instruments was performing structural work that the instruments themselves cannot replicate. As ranking precision rises, the buffer that imprecision provided to certain populations vanishes, and the system converges toward near-total exclusion at the tail. This does not vindicate discretionary harms, which are real and well documented. It does mean that the equity case for high-fidelity algorithmic ranking under scarcity is weaker than the field has assumed.

Our central quantity $D$ is a ratio of selection probabilities. This choice to measure inequality in relative rather than absolute terms is a substantive one that warrants defence. Under rationing, selection probabilities are small for every group, and an equity metric grounded in absolute differences will report small numbers regardless of how the system distributes. The politically and morally salient question is not how often any particular group is selected but how the chance of selection is routed across the population, which is the structural question that Kasy and Abebe \shortcite{kasy2021fairness}
~identify as primary. A ratio captures this routing directly. When $D$ grows from 2 to 10 under increasing fidelity, the lived experience for members of the group disadvantaged by the allocation system is that the system has moved from partially accessible to effectively closed, even though the absolute probability change may look modest. Relative disparity is the quantity that tracks what affected populations actually encounter when they interact with the system. 

Several limitations warrant discussion. Our theoretical model assumes two groups with Gaussian latent distributions, equal within-group variance, and a single shared scalar fidelity parameter $\rho$. Real-world systems involve intersecting group memberships, multidimensional scores, heteroscedastic noise, and group-specific measurement quality, which is typically worse for marginalised populations because of sparser records and fewer prior interactions with the system. Group-specific fidelity $\rho_\text{B} < \rho_\text{A}$ is, by our analysis, expected to deepen rather than mitigate the trap. Thus, we treat the shared-$\rho$ formulation as a lower bound on the disparity real systems exhibit. 
The log-normal check and convergent findings from Moon and Guha's \shortcite{moon2026paradox} discrete combinatorial framework suggests that the qualitative phenomenon is not an artefact of the Gaussian assumptions used here, even if the precise functional form of divergence may differ in more complex settings.

Second, our two empirical validations span distinct countries, populations, and algorithmic architectures, yet both operate within public-sector welfare systems. Whether the Accuracy Trap manifests comparably in market-based allocation, or in settings with moderate rather than extreme scarcity, remains an open empirical question. Third, the mapping from real-world instruments to the theoretical fidelity parameter~$\rho$ involves simplifying assumptions about how scoring systems decompose into signal and noise components. We report this mapping transparently in Section \ref{sec:empirical_validation} and we note that $\rho$ as we operationalise it indexes rank-discrimination rather than calibration, which is the dimension of model performance most relevant to threshold-based selection. Fourth, although we defend relative disparity as the policy-relevant measure, we acknowledge that $D$ becomes statistically unstable at the deepest scarcity regimes where both probabilities in the ratio are small. At $1-\sigma = 0.95$, the synthetic case probabilities are $P_A = 0.074$, $P_B = 0.026$, the SEER probabilities are $P_A = 0.088$, $P_B = 0.045$, and the Child Welfare probabilities are $P_A = 0.091$, $P_B = 0.032$. This demonstrates that the relative measure is not artificially inflated by small denominators (we report the full absolute selection probabilities and $D$ in Supplementary Material Table 3). 

These findings have direct implications for algorithmic governance. Current audit frameworks for public-sector algorithms focus predominantly on bias, asking whether a model's error rates or score distributions differ across protected groups \cite{chouldechova2017fair, hardt2016equality}.
~Our results imply that bias audits are necessary but insufficient. Agencies deploying algorithmic prioritisation under scarcity should additionally audit for \textit{allocative volatility}, which we define as the sensitivity of relative access probabilities to changes in fidelity and supply. Operationally, this means reporting the disparity ratio $D(\sigma, \rho)$ across a range of plausible scarcity levels and fidelity values, as a stress test for equity under budget contraction. Where such audits reveal exponential sensitivity, policymakers face a structural choice. They may introduce intentional imprecision through mechanisms such as banding or weighted lotteries, or they may address the scarcity constraint directly through supply-side expansion. 
Consider, for example, a child welfare agency whose risk scoring tool exhibits high allocative volatility in the region of its actual operating point. Moving from a precise rank-ordering to a banded system in which all cases above a threshold quantile receive equal priority (with tie-breaking by lottery within the band) would reduce the system's ability to distinguish fine-grained risk differences but would simultaneously reduce the exponential amplification of between-group disparity that our analysis predicts. Whether this trade-off is acceptable is a policy question, not a technical one, and our framework makes it visible. Optimising the sorting algorithm alone cannot resolve a disparity whose origin lies in the interaction between precision and scarcity.


\appendix

\section{Formal Derivations of the Accuracy Trap}\label{secA1}

We provide formal derivations for the Accuracy Trap here. 

\subsection{Model Setup and Population Distributions} \label{sec:model_setup}

We model a population partitioned into two groups: Group A (structurally advantaged by the allocation system) and Group B (baseline). Each individual has a latent risk value $Z$. We assume unit variance, defining the structural gap $\Delta$ via group means, $Z_\text{A} \sim \mathcal{N}(\mu_\text{A},\, 1)$ and $Z_\text{B} \sim \mathcal{N}(\mu_\text{B},\, 1)$ such that:

\begin{equation}
\Delta = \mu_\text{A} - \mu_\text{B} > 0
\end{equation}

\noindent The algorithm does not observe $Z$ directly. Instead, it outputs a score $Y$ determined by a fidelity parameter $\rho \in [0,1]$ and independent noise $\varepsilon \sim \mathcal{N}(0,1)$:
\begin{equation}
    Y = \rho Z + \sqrt{1 - \rho^2}\,\varepsilon, \qquad \varepsilon \perp Z
\end{equation}

\noindent\ This construction preserves unit variance for all $\rho$ such that the groupwise observed score distributions are $Y_\text{A} \sim \mathcal{N}(\rho\mu_\text{A},\, 1)$ and $Y_\text{B} \sim \mathcal{N}(\rho\mu_\text{B},\, 1)$.

\subsection{Proposition 1 (Gaussian Regime): Exponential Tail Amplification Under Scarcity} \label{sec:prop_1}

Let $\sigma \in (0,1)$ denote the scarcity level, interpreted as the fraction of the population that can be selected. Let $t = t(\sigma)$ be the policy cutoff such that selection occurs when $Y > t$ and the selected fraction equals $\sigma$, i.e.  $P(Y > t) = \sigma$. We treat $\mu_\text{A}$, $\mu_\text{B}$, and $\rho$ as fixed constants as $\sigma$ varies.

As scarcity increases ($\sigma \to 0$), the cutoff moves into the extreme right tail ($t \to \infty$). For fixed $\rho$ and group means $\mu_\text{g}$, this implies $t - \rho\mu_\text{g} \to \infty$ for each group $g$, justifying the use of tail asymptotics. 

\noindent We define relative disparity as the ratio of selection probabilities:
\begin{equation}
    D(t,\rho) = \frac{P(Y_\text{A} > t)}{P(Y_\text{B} > t)}
\end{equation}

\noindent Because $Y_\text{g} \sim \mathcal{N}(\rho\mu_\text{g}, 1)$, we have:
\begin{equation}
    P(Y_\text{g} > t) = 1 - \Phi(t - \rho\mu_\text{g})
\end{equation}

\noindent For large $x$, Mill's ratio gives \cite{mills_1926} (for background on Mill's ratio see \cite{Vershynin_2026, FROM2020123872}):
\begin{equation}
    1 - \Phi(x) \sim \frac{\phi(x)}{x} \quad \text{as } x \to \infty
\end{equation}

\noindent Applying this with $x_\text{g} = t - \rho\mu_\text{g}$ yields:

\begin{equation}
    P(Y_\text{g} > t) \sim \frac{\phi(t - \rho\mu_\text{g})}{t - \rho\mu_\text{g}}
\end{equation}
\noindent Hence: 
\begin{equation}
    D(t,\rho) \sim \frac{\phi(t - \rho\mu_\text{A})}{\phi(t - \rho\mu_\text{B})} \cdot \frac{t - \rho\mu_\text{B}}{t - \rho\mu_\text{A}}
\end{equation}

\noindent The term $\frac{t - \rho\mu_\text{B}}{t - \rho\mu_\text{A}} \to 1$ as $t \to \infty$, so the dominant asymptotic behavior is governed by the Gaussian density ratio:
\begin{equation}
    \frac{\phi(t - \rho\mu_\text{A})}{\phi(t - \rho\mu_\text{B})} = \exp\!\left(-\tfrac{1}{2}\left[(t - \rho\mu_\text{A})^2 - (t - \rho\mu_\text{B})^2\right]\right)
\end{equation}

\noindent Expanding the quadratic difference:
\begin{align}
    &-\tfrac{1}{2}\left[(t - \rho\mu_\text{A})^2 - (t - \rho\mu_\text{B})^2\right] \notag\\
    &\quad= -\tfrac{1}{2}\Big[(t^2 - 2t\rho\mu_\text{A} + \rho^2\mu_\text{A}^2) - (t^2 - 2t\rho\mu_\text{B} + \rho^2\mu_\text{B}^2)\Big]
\end{align}

\noindent The $t^2$ terms cancel, giving:
\begin{align}
    &-\tfrac{1}{2}\left[-2t\rho\mu_\text{A} + 2t\rho\mu_\text{B} + \rho^2(\mu_\text{A}^2 - \mu_\text{B}^2)\right] \notag\\
    &\quad= t\rho(\mu_\text{A} - \mu_\text{B}) - \tfrac{1}{2}\rho^2(\mu_\text{A}^2 - \mu_\text{B}^2)
\end{align}

\noindent Substituting $\Delta = \mu_\text{A} - \mu_\text{B}$:

\begin{equation}
    = t\rho\Delta - \tfrac{1}{2}\rho^2(\mu_\text{A}^2 - \mu_\text{B}^2)
\end{equation}
\noindent The second term is constant in $t$, while the first grows linearly in $t$. Therefore:
\begin{equation}
    \ln D(t,\rho) = t\rho\Delta + O(1) \quad \text{as } t \to \infty
\end{equation}

\noindent Equivalently:
\begin{equation} \label{eq:D_leadingorder}
    D(t,\rho) = C(\rho,\mu_\text{A},\mu_\text{B}) \cdot \exp(t\rho\Delta) \cdot (1 + o(1)) \quad \text{as } t \to \infty
\end{equation}

\noindent for some finite constant $C(\rho,\mu_\text{A},\mu_\text{B}) > 0$. In particular, as multiplicative factors converge to a constant:
\begin{equation}
    D(t,\rho) \propto \exp(t \cdot \rho \cdot \Delta) \quad \text{as } t \to \infty
\end{equation}

\noindent If the relevant marginal distribution of $Y$ is approximately standard normal, then $t \approx \Phi^{-1}(1 - \sigma)$. In general (e.g., mixtures of shifted normals), we treat $t = t(\sigma)$ as the cutoff achieving pooled selection rate $\sigma$; the derivation only requires that $\sigma \to 0$ implies $t \to \infty$.

\subsection{Tail-Amplified Sensitivity to Fidelity} \label{sec:tail_amplified}

Using the leading-order form $D(t,\rho) \approx C\exp(t\rho\Delta)$ (Equation \ref{eq:D_leadingorder}) and holding the policy cutoff $t$ fixed, the sensitivity of disparity to fidelity is:

\begin{equation} \label{eq:tailampl_derivS15}
    \frac{\partial D}{\partial \rho} \approx t\Delta \cdot D(t,\rho) \approx t\Delta \cdot \exp(t\rho\Delta)
\end{equation}

\noindent Equivalently: 
\begin{equation}
    \frac{\partial}{\partial \rho}\left[\ln D(t,\rho)\right] \approx t\Delta
\end{equation}




\subsection{Proposition 2 (Heavy-Tailed Regime): Polynomial Divergence for Log-Normal Scores}

We now consider a log-normal regime. Let the observed quantity be $X = exp(Z)$ where:

\begin{equation}
Z = \ln(X) \sim N(\mu, 1)
\end{equation}

\noindent A threshold $X > T$ is equivalent to $Z > \tau$ where:

\begin{equation}
\tau = \ln(T)
\end{equation}

\noindent We assume ranking and measurement operate on the log-risk scale (equivalently, noise is multiplicative in $X$), so the Gaussian derivation applies to $Z$. Substituting $\tau = \ln(T)$ into the leading-order Gaussian result yields:

\begin{equation}
D(T, \rho) \propto \exp( \ln(T) \cdot \rho\Delta )
\end{equation}

\noindent Using $\exp(b \ln a) = a^b$:

\begin{equation}
D(T, \rho) \propto T^{\rho\Delta}
\end{equation}

\noindent Thus, in a log-normal (heavy-tailed) regime, relative disparity grows as a power law in the raw threshold $T$, while the monotone interaction remains: increasing scarcity (raising $T$) or increasing fidelity $\rho$ increases relative disparity in the tail under this log-normal shift model.

\bibliography{aaai2027}


\clearpage
\onecolumn
\newpage


\renewcommand{\thefigure}{\arabic{figure}}
\renewcommand{\thetable}{\arabic{table}}
\renewcommand{\theequation}{\arabic{equation}}
\renewcommand{\thepage}{\arabic{page}}
\setcounter{section}{0}
\setcounter{figure}{0}
\setcounter{table}{0}
\setcounter{equation}{0}
\setcounter{page}{1} 


\section*{Supplementary Material}

\section{Additional Information on the Formalization of the Accuracy Trap}
\label{sec:proofs}
This section provides material that characterizes additional components of the Accuracy Trap. This includes further describing and connecting the fidelity parameter to model performance, and presenting a means through which to empirically demonstrate the approximation appropriateness of the $t \to \infty$ behaviour.

\subsection{Relating Fidelity Parameter to Sorting Fidelity (Operational Interpretation)}

The fidelity parameter $\rho$ admits a direct operational interpretation in this signal-plus-noise model. 
Since we characterize an algorithm's output $Y$ as  $Y = \rho Z + \sqrt{1 - \rho^2}\epsilon$ with $\epsilon \perp Z$ and $\text{Var}(Z) = \text{Var}(\epsilon) = 1$, we have:

\begin{equation}
\text{Corr}(Y, Z) = \rho
\end{equation}

\noindent Thus, increasing $\rho$ increases the linear association between the observed score and the latent risk, and reduces the fraction of score variance attributable to noise $(1 - \rho^2)$.

In the equal-variance Gaussian location model used here (i.e., different means, unit variances), the AUC for distinguishing $Z_\text{A}$ from $Z_\text{B}$ using the observed score $Y$ takes the closed form:

\begin{equation}
\text{AUC} = \Phi\left( \frac{\rho\Delta}{\sqrt{2}} \right)
\end{equation}

\noindent and is strictly increasing in $\rho$.

Equivalently, given an empirical AUC measured on a deployed scoring system with known structural gap $\Delta$, the implied fidelity is $\rho = \sqrt{2}\,\Phi^{-1}(\text{AUC})/\Delta$. This provides a direct operational mapping from observable model performance to the theoretical fidelity parameter used in the Accuracy Trap derivation.


\subsection{Fixed-$\sigma$ Policy Cutoffs (When $t$ Depends on $\rho$)}

Earlier in the main text Appendix A.3 Tail-Amplified Sensitivity to Fidelity, we established that the derivative of $D$ with respect to $\rho$ is as follows: 

\begin{equation} \label{eq:tailampl_derivS15b}
    \frac{\partial D}{\partial \rho} \approx t\Delta \cdot D(t,\rho) \approx t\Delta \cdot \exp(t\rho\Delta)
\end{equation}

This holds for a fixed cutoff $t$. In many allocation systems, however, policy fixes the selected fraction $\sigma$ and the cutoff adjusts endogenously as fidelity changes, i.e., $t = t(\sigma, \rho)$ is defined implicitly by $P(Y > t(\sigma, \rho)) = \sigma$ for the pooled population. In that setting, the total derivative of $\ln D$ with respect to $\rho$ is:

\begin{equation}
\frac{d}{d\rho} [\ln D(t(\sigma, \rho), \rho)] = \frac{\partial}{\partial\rho} \ln D(t, \rho) + \frac{\partial}{\partial t} \ln D(t, \rho) \cdot \frac{\partial t}{\partial\rho}
\end{equation}

\noindent The first term is the direct effect of fidelity at a given cutoff; asymptotically it scales as $t\Delta$. The second term captures the fact that the cutoff may shift as $\rho$ changes. The key mechanism persists in scarce regimes because $\sigma \to 0$ implies $t \to \infty$, and tail behavior drives both $\ln D$ and its sensitivity to fidelity to be dominated by the extreme-tail scaling derived from the proposition from main text Appendix A.2 Proposition 1 (Gaussian Regime): Exponential Tail Amplification Under Scarcity. 


\subsection{Finite-T Validation of the Asymptotic Approximation (Numerical Check)} \label{sec:finite_t}

Main text Appendix A.2 Proposition 1 describes the exponential tail amplification of $D$ under scarcity, providing an asymptotic expression valid as $t \to \infty$. To ensure the approximation is accurate in empirically relevant scarcity regimes, one can numerically compare the following:

\begin{itemize}
    \item Exact disparity: $D_{\mathrm{exact}}(t, \rho) = \frac{[1 - \Phi(t - \rho \mu_\text{A})]}{[1 - \Phi(t - \rho \mu_\text{B})]}$
    \item Asymptotic form: $D_{\mathrm{asym}}(t, \rho) = \hat{C}(\rho) \cdot \exp(t\rho\Delta)$
\end{itemize}

\noindent where $\hat{C}(\rho)$ may be estimated by matching at a reference cutoff or by including the terms $\exp(-\frac{1}{2}\rho^2(\mu_\text{A}^2 - \mu_\text{B}^2))$ and $\frac{t - \rho \mu_\text{B}}{t - \rho \mu_\text{A}}$. We present a graphical comparison between asymptotic and exact approximation in Figure \ref{fig:approx_asym}.

\section{Empirical Validation Additional Findings}
\label{sec:extra-results}

\subsection{Empirical Validation Risk Score Descriptive Statistics}\label{secA2}

This section presents further information on risk scores computed for the cancer ($R_\text{bcss}$) and child welfare ($R_\text{llm}$) empirical validation step. Tables \ref{tab:childwelfare_stats} and \ref{tab:seer_stats} report descriptive statistics for the LLM-derived child welfare risk scores ($R_\text{llm}$) and the SEER-derived breast cancer mortality risk scores ($R_\text{bcss}$), grouped by the structural contrast used in each empirical validation. The tables also report standardized effect sizes (Cohen's d) for the between-group differences in mean risk. These descriptive statistics establish the empirical $\Delta$ values reported in the main text, and confirm that the structural gap on which the Accuracy Trap operates is detectable in both validation datasets prior to any disparity analysis.

\begin{table}[h]
\centering
\renewcommand{\arraystretch}{1.2}
\begin{tabular}{lcc}
\hline
& \multicolumn{2}{c}{Risk score $R_{\text{llm}}$} \\
Statistic & Inner Toronto & Outer Toronto \\
\hline
$N$        & 178    & 405    \\
Mean       & 0.443  & 0.385  \\
Variance   & 0.0549 & 0.0476 \\
Median     & 0.404  & 0.370  \\
Mode       & 0.500  & 0.000  \\
\hline
Cohen's $d$ (95\% CI) & \multicolumn{2}{c}{0.261 \; [0.084, 0.441]} \\
\hline
\end{tabular}
\caption{Statistics on LLM-generated model-child welfare case risk scores ($R_\text{llm}$) grouped by geographic region (Inner and Outer Toronto) and Cohen's D for risk scores between groups.}
\label{tab:childwelfare_stats}
\end{table}

\begin{table}[h]
\centering
\footnotesize
\renewcommand{\arraystretch}{1.05}
\setlength{\tabcolsep}{4pt}
\begin{tabular}{lcccc}
\toprule
& \multicolumn{2}{c}{BCSS outcome} & \multicolumn{2}{c}{Risk score $R_{\text{bcss}}$} \\
\cmidrule(lr){2-3} \cmidrule(lr){4-5}
Statistic & Black & White & Black & White \\
\midrule
BCSS = $1$     & 1,796  & 6,481   & --- & --- \\
BCSS = $0$ & 15,024 & 112,181 & --- & --- \\
Mean              & 0.1068 & 0.0546  & 0.418 & 0.325 \\
Variance          & 0.0954 & 0.0516  & 0.0722 & 0.0595 \\
Median            & --- & --- & 0.374 & 0.236 \\
Mode              & --- & --- & 0.141 & 0.138 \\
\midrule
Cohen's $d$ & \multicolumn{2}{c}{0.218 [0.199, 0.239]} & \multicolumn{2}{c}{0.375 [0.358, 0.392]} \\
\bottomrule
\end{tabular}
\caption{Statistics on SEER breast cancer–specific survival (BCSS) outcome and model-predicted risk scores ($R_{\text{bcss}}$) grouped by race, and Cohen's $d$ of risk scores between groups. 95\% CIs shown in brackets.}
\label{tab:seer_stats}
\end{table}

\subsection{Additional Simulation-Based Empirical Validation of the Accuracy Trap} 

\begin{figure}[H]
\centering
\includegraphics[width=0.7\textwidth]{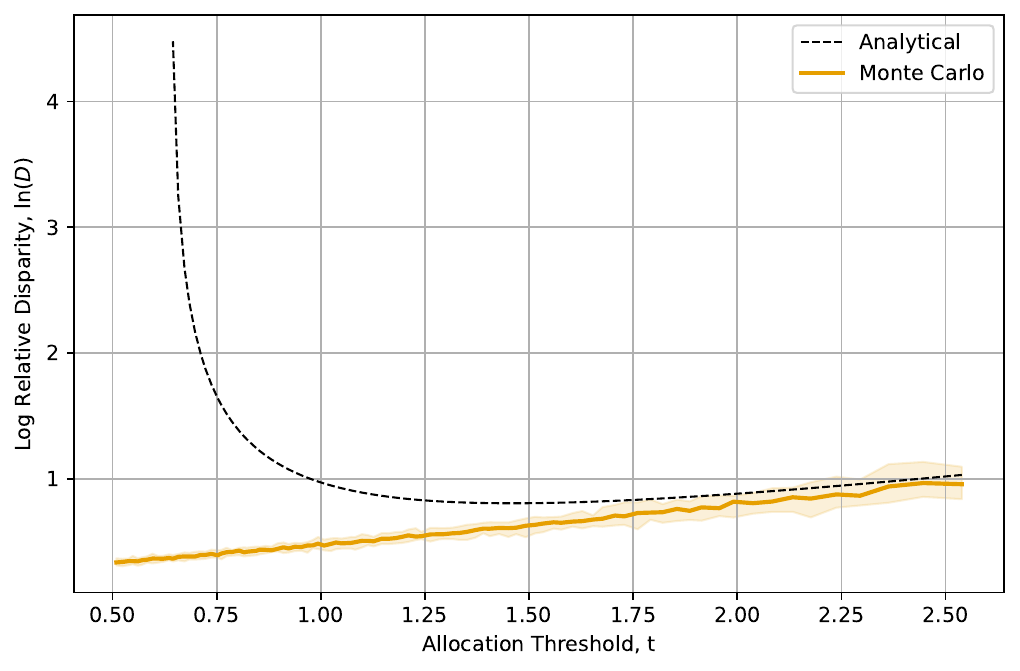}
\caption{Convergence of Monte Carlo simulation to the analytical simulation under high resource scarcity. The log relative disparity in Monte Carlo simulation converges to the analytical solution as allocation threshold (scarcity) increases. Simulation were performed with two Gaussian distributions with $\mu_\text{A} = 0.8$, $\mu_\text{B} = 0.3$, $\text{SD} = 1$, and $\rho = 0.8$. Results are shown for $t > 0.5$ to demonstrate convergence in the tail region where Mill's ratio applies. Shaded areas represents 95\% confidence intervals across 25 simulation iterations. This figure operationalizes the finite-T validation procedure described above in Section \ref{sec:finite_t} of the Supplementary Material.}
\label{fig:approx_asym}
\end{figure}

\begin{figure}[H]
\centering
\includegraphics[width=0.7\textwidth]{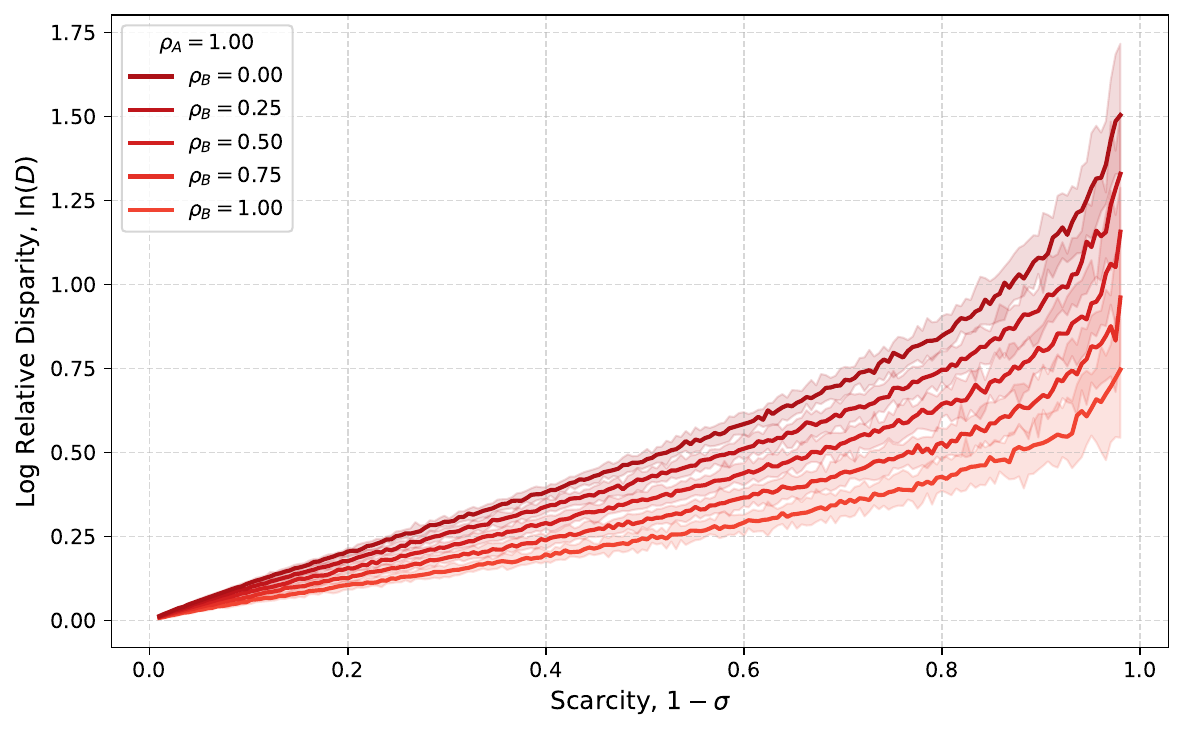}
\caption{We examine how reducing the fidelity of one group while holding the fidelity of the other group fixed impacts the log relative disparity as scarcity increases. We model two Gaussian distributions, both with $\text{SD} = 1$, with $\mu_\text{A} = 0.6$ and $\mu_\text{B} = 0.3$. Lines depict median log relative disparity and shaded regions denote $95\%$ confidence intervals derived from 25 Monte Carlo iterations. As $\rho_\text{B}$ decreases (but $\rho_\text{A}$ remains fixed), the observed mean for Group B observed scores ($Y_B$) shifts towards zero, increasing the separation between the two groups. This figure provides further explorations of the Accuracy Trap as we examine the impact of asymmetric group fidelity on log relative disparity. }
\label{fig:s3_reducingfidelity}
\end{figure}

\begin{figure}[H]
\centering
\includegraphics[width=\textwidth]{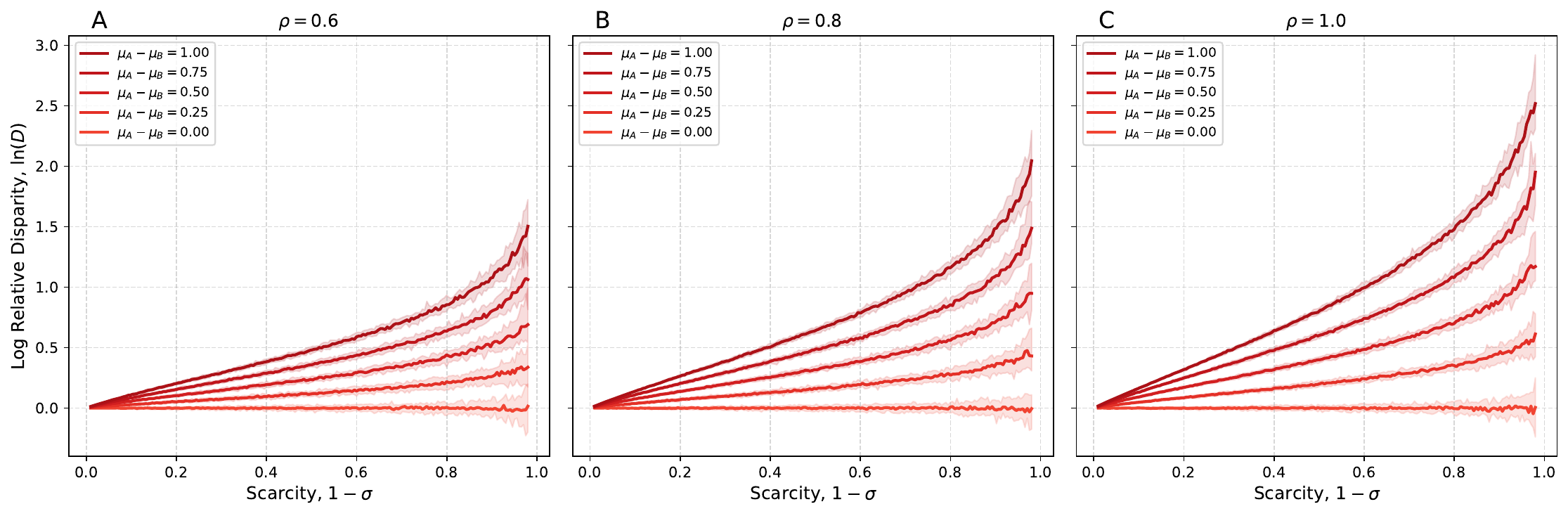}
\caption{Amplification of log relative disparity as a function of structural gap, scarcity, and fidelity. Increasing the difference in means ($\mu_\text{A}-\mu_\text{B}$) between two groups amplifies relative disparity under high scarcity ($1-\sigma \to1)$ and fidelity ($\rho$). Plots depict simulations for two Gaussian distributions, $A$ and $B$ with $\text{SD}=1$. Group $B$ is held fixed at $\mu_\text{B} = 0.3$  while $\mu_\text{A}$ varies such that the structural gap ($\mu_\text{A} - \mu_\text{B}$) ranges from 0 to 1 in the figures. Panels show results for three fidelity levels: (A) $\rho = 0.6$, (B) $\rho = 0.8$, and (C) $\rho = 1.0$. This figure supports our main text, Empirical Validation Results, showing that as $\rho$ increases (moving from the left panel to the right), the allocation algorithm better reflects the underlying structural gap $\Delta$ between groups, leading to greater relative disparities in high-scarcity areas.}
\label{fig:s2_changingdiffmeans}
\end{figure}

\begin{figure}[H]
\centering
\includegraphics[width=\textwidth]{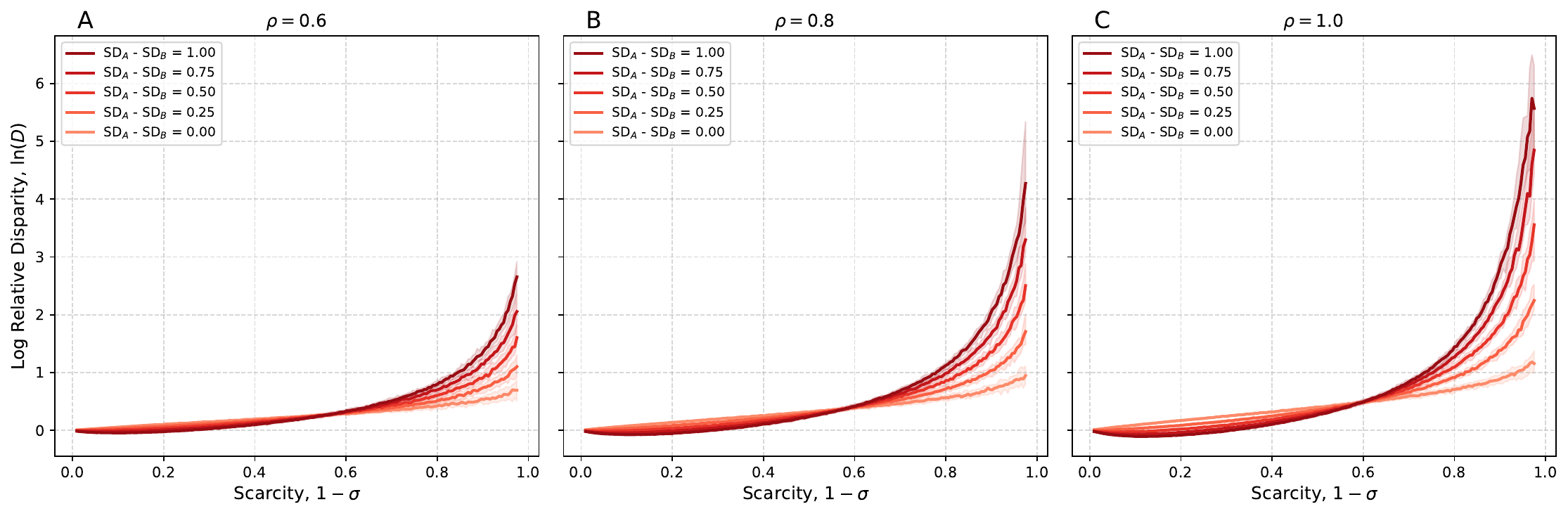}
\caption{We examine how increasing the difference in variances between groups $A$ and $B$ impacts relative disparity as scarcity increases. We model two Gaussian distributions: $\text{SD}_\text{B} = 1$ is fixed and $\text{SD}_\text{A}$ varies such that $\text{SD}_\text{A} - \text{SD}_\text{B}$ ranges from 0 to 1, with $\mu_\text{A} = 0.8$ and $\mu_\text{B} = 0.3$. Panels show results across three fidelity levels where both distributions have the same fidelity: (A) $\rho = 0.6$, (B) $\rho = 0.8$, and (C) $\rho = 1.0$. Higher variance in Group A can lead to lower relative disparity at low scarcity. However, under high scarcity ($1-\sigma \to 1$), we observe non-linear amplification of disparity. Lines depict median log relative disparity and shaded regions denote $95\%$ confidence intervals derived from 25 Monte Carlo iterations. This figure provides further explorations of the Accuracy Trap as we examine the impact of asymmetric group variances on log relative disparity. }
\label{fig:s4_diff_variances}\end{figure}

Figures \ref{fig:approx_asym}-\ref{fig:s4_diff_variances} presents further explorations of the Accuracy Trap following Monte Carlo simulation

\subsection{Robustness of Relative Disparity to Small-Denominator Instability}

The primary metric of our analysis is relative disparity $D$. When both group allocation probabilities in the ratio are small, $D$ becomes statistically unstable. Small allocation probabilities occur especially in regimes of high scarcity. To demonstrate that our Empirical Validations findings' growing disparity results are not artificially inflated by the metric's instability with small denominators, the median allocation probabilities for each group are provided in Table 3. Results are shown for full fidelity ($\rho = 1$), the regime under which disparity is most pronounced. Table \ref{tab:allocation_disparity} shows that each median group allocation probability is considerably above zero across all gradations of scarcity, even under high scarcity. 


\begin{table*}[h]
\centering
\begin{tabular}{ccccccccccc}
\toprule
& \multicolumn{3}{c}{Synthetic} & \multicolumn{3}{c}{SEER} & \multicolumn{3}{c}{Child Welfare} \\
\cmidrule(lr){2-4} \cmidrule(lr){5-7} \cmidrule(lr){8-10}
Scarcity $(1-\sigma)$ & $P_A$ & $P_B$ & $\ln(D)$ & $P_A$ & $P_B$ & $\ln(D)$ & $P_A$ & $P_B$ & $\ln(D)$ \\
\midrule
0.05 & 0.974 & 0.926 & 0.051 & 0.968 & 0.948 & 0.021 & 0.970 & 0.939 & 0.033 \\
0.15 & 0.906 & 0.794 & 0.132 & 0.903 & 0.843 & 0.069 & 0.876 & 0.837 & 0.046 \\
0.25 & 0.829 & 0.671 & 0.211 & 0.831 & 0.739 & 0.118 & 0.783 & 0.735 & 0.064 \\
0.35 & 0.741 & 0.559 & 0.281 & 0.754 & 0.635 & 0.172 & 0.716 & 0.618 & 0.147 \\
0.45 & 0.648 & 0.452 & 0.358 & 0.673 & 0.533 & 0.234 & 0.606 & 0.524 & 0.146 \\
0.55 & 0.548 & 0.352 & 0.442 & 0.587 & 0.431 & 0.310 & 0.497 & 0.429 & 0.148 \\
0.65 & 0.442 & 0.258 & 0.537 & 0.486 & 0.331 & 0.386 & 0.407 & 0.325 & 0.224 \\
0.75 & 0.329 & 0.171 & 0.651 & 0.373 & 0.233 & 0.472 & 0.326 & 0.215 & 0.416 \\
0.85 & 0.207 & 0.093 & 0.800 & 0.237 & 0.138 & 0.543 & 0.201 & 0.126 & 0.465 \\
0.95 & 0.074 & 0.026 & 1.07 & 0.088 & 0.045 & 0.683 & 0.091 & 0.032 & 1.050 \\
\bottomrule
\end{tabular}
\caption{Median group resource allocation rates ($P_A$, $P_B$) and log relative disparity ($ln(D)$) across different scarcity levels ($1-\sigma$) when $\rho=1$ for the three domains: (i) Synthetic Gaussian data Group A ($\mu_A=0.8, SD=1$) and Group B ($\mu_B=0.3, SD=1$); (ii) SEER breast cancer mortality where Group A is Non-Hispanic Black and Group B is Non-Hispanic White; (iii) Child welfare where Group A represents Inner Toronto teams and Group B represents Outer Toronto teams. In all domains, Group A corresponds to the group with a higher mean risk. We report median allocation rates across 25 independent Monte Carlo simulations for the synthetic and SEER domains, and 1,000 bootstrap iterations for the child welfare domain. This table finds support that relative disparities are not artificially inflated through small denominators.}
\label{tab:allocation_disparity}
\end{table*}

\section{Dataset Information and Model Training Details}

\subsection{LLM Child Welfare Case Note Classification Details}

The child welfare data for this paper included two types of anonymised narrative documents for families engaged with the child welfare agency: Regular Case notes and Service Plan documents.

Regular case notes contain narrative records of a family’s case with the agency, including interactions, observations, and correspondence among parties involved in the case (e.g., families, service providers such as doctors and lawyers). Each Regular Case note also includes metadata, including a family’s reference number, mode of interaction, the time, date, and location where the documented event occurred. These Regular Case notes are essential records that track child welfare progress and can serve as evidence in legal proceedings.

Service Plan documents record actionable goals that family members need to work on to successfully address child safety concerns raised by the agency and close their case with the agency. A single Plan may have multiple goals. This is an important document that also helps guide caseworkers in supporting families. In addition to goals for the family, the Plan includes metadata, including the family’s reference number and the time period during which it is applicable. A Plan is valid for 6 months, and a new Plan is drawn up if the family remains engaged with the agency.

Service Plan documents are closely tied with Regular case notes as records of progress made towards goals outlined in the Plan are tracked and recorded in the Regular case notes. Thus, to establish if child welfare case notes documented progress toward case goals outlined in the Service Plan, we followed the methodology by Moon et al. \shortcite{moon2026promises}, who previously examined the feasibility of employing Local LLMs to track case goal progress on this dataset. We used a local instance of Meta-Llama-3.1-8B for zero-shot classification. To ensure data privacy and protection, we employed a local LLM. The Local LLM was provided with case note narrative text and applicable case goal (i.e., activity) and given the following prompt with temperature = 0.1. 

\begin{list}{}{
    \leftmargin=2em 
    \rightmargin=2em 
}
\item[] 
\small
\textbf{Prompt:} 

\begin{enumerate}
    \item[] \texttt{You are analyzing a child welfare worker reviewing case notes. Do the following:}
    \item \texttt{Read through the casenote and store it in summary.}
    \item \texttt{Assess whether the summary indicates progress toward completing the following activity: \{\detokenize{activity_name}\}. \\Answer strictly `Yes' or `No'.}
    \item[] \texttt{Case Note: \{\detokenize{narrative_text}\}}
\end{enumerate}
\end{list}

The model was instructed to output a single binary label (1 = aligned with case goals, 0 = not aligned). Prior work by Moon et al. \shortcite{moon2026promises} performed validation of this task against a manually labeled subset of n=6,031 case notes written for 100~families, randomly stratified by case duration. 

To conduct a disparity analysis between cases served by Inner and Outer Toronto teams, we merged the LLM-labeled case notes with metadata provided by the agency. This metadata included case reference numbers, postal codes for primary caregivers, and the specific geographic team assigned to each family. We also used the metadata to visualize geographic disparity in Figure 1 of the main paper. 



\subsection{SEER Gradient Boosting Model Training Details}

We trained a gradient boosting classifier to predict 5-year breast cancer-specific survival (BCSS) using the SEER registry. The implementation used \textsc{scikit-learn}~1.6.1 GradientBoostingClassifier. The feature set used to train the model included age at diagnosis, AJCC tumour stage (7th ed.), tumor grade, tumor size, ER status, PR status, HER2 status, number of regional nodes examined, and number of regional nodes positive \cite{seer_descript}. Categorical variables, stage and grade, were converted to ordinal integers 1-4 to reflect the four graded levels of severity corresponding to the variables. Missing values were handled via complete-case analysis. After applying all exclusion criteria, 24.3\% of Black patients and 27.7\% of White patients were retained in the pre-processed dataset. The comparable exclusion rates suggest processing had a similar effect across groups.

The pre-processed dataset (N = 135,482) was split into training (80\%) and test (20\%) partitions stratified jointly by BCSS outcome and race. Hyperparameters were tuned via RandomizedSearchCV with 100 iterations, 5-fold cross-validation on the training set, and scored by AUC. The range of hyperparameters in the search were: max\_depth = (3, 7), learning\_rate = (0.01, 0.20), n\_estimators = (100, 500), subsample = (0.4, 1.0), min\_samples\_split = (2, 20), and min\_samples\_leaf = (1, 10). 

The final model used max\_depth = 4, learning\_rate = 0.04071, n\_estimators = 211, subsample = 0.6406, min\_samples\_split = 4, and min\_samples\_leaf = 8, with balanced class weights to address the \~6\% rate of breast-cancer specific death within 5 years of diagnosis. The model achieved a held-out AUC of 0.8471 (95\% CI [0.8370, 0.8564]) on the test set, with CI estimated via 1,000 bootstrap samples. Race was excluded from the feature set during training to ensure that any disparity observed in the empirical validation reflects the structural interaction of fidelity and scarcity rather than direct race-based prediction.

\section{Implementation and Reproducibility Details}
\label{sec:repro}

\subsection{Dataset Usage}
\subsubsection{Dataset Motivation}
One of the primary goals with the empirical validation was to demonstrate the prevalence of the Accuracy Trap across different contexts. As such, two datasets of distinct nature were selected. The SEER registry was selected since it is a widely used publicly accessible dataset on U.S. cancer outcomes, a domain in which model-predicted risk and resource allocation are already salient areas of interest. By validating our findings with this dataset, we are able to demonstrate that the Accuracy Trap exists in the domain of healthcare. The child welfare dataset was used to demonstrate the Accuracy Trap's presence in a different algorithmic context and domain, namely child welfare, and in a particular context where risk scores are generated with LLMs as opposed to traditional ML like the gradient-boosted classification model we trained and used on the SEER registry data. 


\subsubsection{Dataset Availability}

The SEER dataset is publicly available: https://seer.cancer.gov \cite{seer2024registry} and can be accessed through a Research Data Use Agreement. Although this data is accessible to researchers, individuals must sign a specific Data-Use Agreement and complete an online application process to gain access. We cannot distribute the data ourselves. 

The Child Welfare dataset is bound by a data-sharing agreement with a government agency to protect the privacy of vulnerable populations. Under our research agreement with the agency, we are prohibited from publicly sharing or posting the data. Authors who would like to replicate our study may contact the corresponding author to initiate a formal application process for access to the data. 


%

\subsection{Computational Experiments}

\subsubsection{Hyperparameters}
The SEER Gradient Boosting Model used in empirical validation required hyperparameter tuning. Details on the hyperparameter settings including the range of values tried and the  criterion used for selection is   provided in C.2 



\subsubsection{Randomization Seed}
A random seed value of 42 was used for all components of simulation, model training, and bootstrapping within the empirical validation code. 

\subsubsection{Infrastructure \& Software}
Experiments were run with a MacBook Pro with a Apple M3 Pro chip,  18 GB memory, and macOS Sequoia 15.5. The software libraries and corresponding versions used are: numpy = 1.24, pandas = 2.0, scipy = 1.11, scikit-learn = 1.3, matplotlib = 3.7, seaborn = 0.12, geopandas = 0.14, shapely = 2.0, pyproj = 3.6, and tqdm = 4.65. 

\subsubsection{Evaluation Metric Selection}
The primary metric used in experiments was relative disparity $D$ between allocation probabilities of two groups. This metric was selected because they are able to reflect the phenomena of resource lockout that occurs when allocation systems structurally disadvantage a particular group such that it is as if the resource is inaccessible to them. An absolute difference disparity metric would be unable to capture this as effectively. For a detailed discussion of this choice of metric, see Section 4. Discussion of the main text. 

\subsubsection{Number of Algorithmic Runs}
The number of runs for each of the empirical validation experiments are provided in Section 3 Empirical Validation of the main text. For each presented result within the figures and tables of the main and supplementary text, the results were generated with 25 independent iterations for the Monte Carlo and SEER domains, and 1,000 bootstrap iterations for the Child Welfare domain. 


\end{document}